\documentclass[12pt,preprint]{emulateapj}
\usepackage{amsmath}

\newcommand{\beq}{\begin{equation}}
\newcommand{\eeq}{\end{equation}}

\def\H2{{{\rm H}_2}}
\def\HI{{\rm H\,I}}
\def\HII{{\rm H\,II}}

\def\dim#1{\mbox{\,#1}}
\def\msunpc2{\dim{M$_\odot$/pc$^2$}}

\def\figname#1#2{#2}

\begin{document}

\title{Modeling Molecular Hydrogen and Star Formation in
  Cosmological Simulations}

\author{Nickolay Y. Gnedin\altaffilmark{1,2,3}, Konstantinos
  Tassis\altaffilmark{2,3,5},  \& Andrey V.  
Kravtsov\altaffilmark{2,3,4}}
\altaffiltext{1}{Particle Astrophysics Center, Fermilab, Batavia, IL 60510}
\altaffiltext{2}{Department of Astronomy and Astrophysics, 
The University of Chicago, Chicago, IL 60637}
\altaffiltext{3}{The Kavli Institute for Cosmological Physics, 
The University of Chicago,
Chicago, IL 60637}
\altaffiltext{4}{Enrico Fermi Institute, The University of Chicago,
Chicago, IL 60637}
\altaffiltext{5}{Current address: Jet Propulsion Laboratory, California Institute of Technology, Pasadena, CA 91109}

\begin{abstract}
We describe a phenomenological model for molecular hydrogen formation
suited for applications in galaxy formation simulations, which
includes non-equilibrium formation of H$_2$ on dust and approximate
treatment of both its self-shielding and shielding by dust from the
dissociating UV radiation.  The model is applicable in simulations in
which individual star forming regions -- the giant molecular complexes
-- can be identified (resolution of tens of pc) and their mean
internal density estimated reliably, even if internal structure is not
resolved. In agreement with previous studies, calculations based on our
model show that the transition from atomic to fully molecular phase depends
primarily on the metallicity, which we assume is directly related to
the dust abundance, and clumpiness of the interstellar medium.  The
clumpiness simply boosts the formation rate of molecular hydrogen,
while dust serves both as a catalyst of H$_2$ formation and as an
additional shielding from dissociating UV radiation. The upshot is
that it is difficult to form fully-shielded giant molecular clouds
while gas metallicity is low. However, once the gas is enriched 
to $Z\sim 0.01-0.1Z_{\odot}$, the subsequent star formation and
enrichment can proceed at a much faster rate.  This may keep star
formation efficiency in the low-mass, low-metallicity progenitors of
galaxies very low for a certain period of time with the effect similar
to a strong ``feedback'' mechanism. The effect may help explain the
steep increase of the mass-to-light ratio towards smaller masses 
observed in the local galaxy population. We apply the model and star
formation recipes based on the local amount of molecular gas to an
output from a cosmological simulation of galaxy formation and show
that resulting global correlations between star formation and gas and
H$_2$ surface densities are in good agreement with observations.
\end{abstract}


\maketitle

\section{Introduction}
\label{sec:intro}

Detailed understanding of galaxy formation remains one of the most
important goals of modern astrophysics. A critical challenge for
achieving this goal, in addition to a sheer variety of physical
processes and the range of spatial and temporal scales involved, is our poor
understanding of how gas is converted into stars under different
conditions. This conversion controls critical aspects of the final
galaxy properties from its total stellar mass and star formation
history to its morphology, which depends on the relative timing of gas
conversion to stars and the epoch of major mergers and rapid mass
assembly.

Recent observational and theoretical progress in understanding
 star formation on the scale of molecular clouds points towards a 
prescription for star formation on larger, galactic scales. In
particular, the observational data support the theoretically
motivated view of molecular clouds in which the specific
star formation rate per local free fall time in the molecular gas is
essentially independent of the local gas density
\citep{sfr:km05,sfr:kt07,sfr:kmt08}. This is fortunate because 
the efficiency of star formation can then be modeled in a simulation
as long as (i) one can resolve and identify the regions corresponding
to star forming molecular clouds in a simulation and (ii) achieve 
resolution sufficient to model their mean internal density properly.

Until recently, these conditions were not easily achievable in
cosmological simulations due to numerical limitations on spatial and
mass resolution. The standard approach in galaxy formation modeling so
far \citep[c.f. recent studies by][and many earlier works, comprehensively reviewed in
the last
reference]{sims:bgbw07,sims:gwmb07,sims:ck07,sims:stws08,sims:cdsh08,sims:od08,sims:tb08,sims:mgk08}
was to adopt a recipe which ties the local star formation rate density
to the gas density via a universal relation, often with threshold
conditions on the gas properties (density, temperature, etc). Such
recipes are loosely based on the empirical correlations observed for
local galaxies \citep[the Kennicutt-Schmidt, hereafter KS,
relations:][]{schmidt59,schmidt63,kennicutt89,sfr:k98a}.  However,
these relations have only been studied relatively well for 
 nearby massive or star bursting galaxies. There is, however, a
growing body of evidence that the relations for galaxies of lower
surface brightness and/or metallicity may be quite different
\citep[see, e.g.,][for references and
discussion and the recent study by \citeauthor{bigiel_etal08} \citeyear{bigiel_etal08}]{robertson_kravtsov08}.

Indeed, as we emphasize below \citep[see
also][]{elmegreen89,elmegreen93,schaye01,pelupessy_etal06,krumholz_etal08}, the transition from
atomic to fully molecular gas depends strongly on the local gas
metallicity and dissociating UV flux. Thus, for example, we expect
that in low metallicity galaxies  gas becomes fully molecular
(and thereby conducive to star formation) at higher
gas density compared to more massive, higher metallicity systems. This
means that on the global scale of a galaxy gas may be converted into stars
at a slower rate per unit mass of gas or, in other words, the star
formation efficiency is lower, even if the density distribution of ISM
is the same. Not only is this regime applicable to
dwarf or low surface brightness galaxies today, but also to the
majority of progenitors of massive galaxies at high redshifts.

The lower efficiency of gas conversion into stars at high redshifts
may have profound implications for galaxy evolution from shaping the
faint end of the luminosity function to the morphological mix of
galaxies. Traditionally, in cosmological simulations and semi-analytic
models the efficiency of star formation in $\lesssim L_{\ast}$
galaxies is suppressed by supernova feedback. However, the efficiency
of such suppression is not clear and in many observed dwarf galaxies
star formation is inefficient without obvious signs of active feedback. 
We argue that a similar suppression can be provided by the 
inherent difficulty of building self-shielding molecular regions 
within low-density, low-metallicity ISM of smaller galaxies. 

Current self-consistent cosmological simulations of galaxy formation
can resolve scales of molecular clouds \citep[i.e., tens of
parsecs,][]{sims:k03,sims:ck07,sims:tb08,sims:r08,sims:gcst08} and
can, therefore, explore the effect of ISM process on the efficiency
of star formation self-consistently. Observational evidence for the universal 
efficiency of gas conversion per free fall time in molecular clouds 
can then be adopted as a well-motivated and robust star formation 
prescription, even if the internal density structure of the star forming
regions is not well resolved. Such recipe is promising but requires 
a model for the formation and destruction of molecular gas in the simulations
in order to correctly identify the regions of current star formation.

Several such models have recently been developed and implemented in
the SPH simulations
\citep{pelupessy_etal06,booth_etal07,robertson_kravtsov08}.
\citet{pelupessy_etal06} have followed reactions of H$_2$ formation
and destruction, following reactions of H$_2$ formation on dust grains
and dust and self-shielding from UV radiation and using a sub-grid
model for gas clouds, which uses observed cloud scaling relations. One
of the main conclusions of their study was the strong dependence of
the molecular gas fraction on the ambient metallicity of ISM.
\citet{booth_etal07} have adopted a different approach, in which they
model formation of molecular clouds using a model based on collisions
between clouds representing molecular clouds as ballistic particles
coagulating upon collision. They have demonstrated that with such a
model many observed properties of disc galaxies can be reproduced.
Recently, \citet{robertson_kravtsov08} presented a model of star
formation based on the local molecular gas content. The latter was
modeled by using pre-computed tables of molecular fraction as a
function of local gas properties (density, temperature, metallicity
and UV flux). The authors showed that the model implied a significant
steepening of the Kennicutt-Schmidt relations in low surface density,
low molecular fraction environments characteristic of dwarf and low
surface brightness galaxies, as well as of the outskirts of disks in
larger galaxies. This study has indicated that one of the key factors
controlling the amount of diffuse molecular gas in the lower density ISM
is UV radiation from the recent local star formation.

In this paper we present a model of molecular gas evolution
and H$_2$-based star formation together with their implementation in the Adaptive
Refinement Tree (ART) code for cosmological simulations. The ART code
is a Eulerian hydrodynamics$+N$-body code, which uses Adaptive Mesh Refinement
(AMR) technique to reach high resolution in the regions of interest (i.e., in
high-density regions of ISM in the case of this study). We present details
of the model of molecular hydrogen formation and disruption and test
the model against available observations. We discuss several star formation
recipes for the gas conversion in molecular clouds and present 
resulting Kennicutt relations, when such recipes and the molecular hydrogen
model are applied to a realistic cosmological simulation of a Milky Way progenitor. 
In a follow-up study, we plan to explore the differences of such model
compared to the old star formation prescriptions in full cosmological 
simulations.

\section{Method}\label{sims}

\subsection{Description of Simulations}
\label{dsims}

The cosmological simulation used as a testing ground for 
our model for molecular hydrogen formation has been described in
detail in \citet{tkg08} as a run ``FNEC-RT''. Here we only recap
 that the simulation was performed with the Eulerian,
gas dynamics + $N$-body Adaptive Refinement Tree (ART) code
\citep{ketal97,Kth,kravtsov_etal02}, which uses adaptive mesh refinement (AMR) in both the
gas dynamics and gravity calculations to achieve a needed dynamic
range.

The simulation follows a Lagrangian region corresponding to five
virial radii of a system, which evolves to approximately Milky Way 
mass ($M\approx 10^{12}{\rm\ M_{\odot}}$) at $z=0$. The mass
resolution in dark matter of $m_{\rm dm}=1.3\times 10^6\dim{M}_\odot$
(and corresponding resolution in gas dynamics of $m_{\rm
gas}=2.2\times 10^5\dim{M}_\odot$). This Lagrangian region is embedded
into a cubic volume of $6h^{-1}$ comoving Mpc on a side, which is
resolved to a lesser extent throughout the simulation.

The simulation includes star formation and supernova enrichment and
energy feedback, and uses self-consistent 3-D radiative transfer of UV
radiation from individual stellar particles using the OTVET
approximation \citep{ga01}. The simulation follows non-equilibrium
chemical network of hydrogen and helium and non-equilibrium cooling and
heating rates, which make use of the local abundance of atomic,
molecular, and ionic species and UV intensity, followed
self-consistently during the course of the simulation. We use the $z=4$ output
of the simulation as at this epoch the gaseous disk of the main galaxy
is relatively quiescent and has not experienced major mergers since
$z\approx 6$.

A new ingredient in the simulation, beyond what has been described in
\citet{tkg08}, is an empirical model for formation and shielding of
molecular hydrogen on the interstellar dust, which is described in the
following section. 

\subsection{The model for tracking molecular hydrogen}
\label{sec:H2model}

The two key processes controlling the abundance of molecular gas
in the ISM enriched with metals are the formation of $\H2$ on dust grains 
and its shielding from photodissociation by the interstellar radiation field 
by itself (self-shielding) and by dust \citep[see][ for a recent review
and references to a large body of prior work]{gm07}.

Unfortunately, the exact treatment of shielding cannot be achieved
in modern cosmological simulations, as it requires
three-dimensional radiative transfer, including radiative transfer in
the Lyman-Werner bands. Therefore, we adopt a phenomenological treatment of
molecular hydrogen shielding, which is tuned to reproduce available 
observations (see below).

In the optically thin regime (no shielding) and in the absence of
dust, equations for the balance of neutral (atomic) hydrogen and
molecular hydrogen can be written as
\begin{eqnarray}
  \dot{X}_{\HI} & = & R(T)n_e X_{\HII} - X_{\HI}\Gamma_{\HI} -
  2\dot{X}_{\H2}, \nonumber \\
  \dot{X}_{\H2} & = & \dot{X}_{\H2}^{\rm gp},
\label{eq:hbal}
\end{eqnarray}
where $X_i$ denotes a number fraction of baryons
in species $i$,
\[
  X_i \equiv n_i/n_b,
\]
and $n_b$ is the total number density of baryons. In equations
(\ref{eq:hbal}) $R(T)$ is the recombination rate, $n_e$
  is the number 
density of free electrons, $\Gamma_{HI}$ is the hydrogen ionization
rate, and the term $\dot{X}_{\H2}^{\rm gp}$ includes multiple
processes of formation and destruction of molecular hydrogen in the
gas phase reactions, via ${\rm H}^-$ and $\H2^+$ ions. The latter
reactions are comprehensively summarized elsewhere
\citep{sk87,aazn97,gp98,gj07,ga08} and are not important for our
purposes; we keep them in the equations to ensure that our model also
works in the metal-free regime.

Physically, the effect of the geometric shielding by dust grains 
and $\H2$ self-shielding
can be described by effective shielding factors, $S_d$ and $S_\H2$,
which take values from 0 to 1. Incorporating these factors, and adding
molecular hydrogen formation on dust, equations (\ref{eq:hbal}) become
\begin{eqnarray}
  \dot{X}_{\HI} & = & R(T)n_e X_{\HII} - S_d X_{\HI}\Gamma_{\HI}  -
  2\dot{X}_{\H2}, \nonumber \\
  \dot{X}_{\H2} & = & S_d S_\H2 \dot{X}_{\H2}^{\rm gp} + R_d n_b X_\HI
  \left(X_\HI+2X_\H2\right),
  \label{eq:mod}
\end{eqnarray}
where we assume that the dust absorption also affects atomic hydrogen
ionization rate. In principle, this is not a crucial assumption since
one would expect that hydrogen is predominantly neutral in regions
where dust absorption is important. In practice, however, the OTVET
approximation for modeling radiative transfer, which we use in our
simulations, is somewhat
diffusive. In simulations presented here, this diffusion leads to
ionizing radiation leaking into neutral, self-shielded gas. The
numerical diffusion effect is not significant, but introducing a
factor $S_d$ in equation (\ref{eq:mod}a) allows to essentially
eliminate it altogether. In a completely analogous manner, we also
reduce the photo-heating rate of the gas in the evolution equation for
the gas internal energy.

The specific form of the shielding
factors $S_d$ and $S_\H2$ cannot
be computed without a detailed treatment of radiative transfer in the
Lyman-Werner bands. Our goal, therefore, is to find physically
plausible functional forms that are motivated by the observational
data. To this end, we follow \citet[][see also \citeauthor{gm07} \citeyear{gm07}]{db96}  and define these two factors
as 
\[
  S_d = e^{\displaystyle-\sigma_{d,\rm eff}(N_\HI+2N_\H2)}
\]
and
\[
  S_\H2 = \frac{1-\omega_\H2}{(1+x)^2} +
  \frac{\omega_\H2}{(1+x)^{1/2}} e^{\displaystyle-0.00085(1+x)^{1/2}},
\]
where $x\equiv N_\H2/(5\times10^{14}\dim{cm}^2)$ and $\sigma_{d,\rm
  eff}$ and $\omega_\H2$ are treated as adjustable parameters. We find
that we obtain the best fit with the observational data (as we
  demonstrate below) for values of
these parameters $\sigma_{d,\rm eff} =
  4\times10^{-21}\dim{cm}^2$ and $\omega_\H2 = 0.2$, which are
  slightly different from those adopted by \citet{gm07}.

The formation rate coefficient of molecular hydrogen on dust, $R_d$,
has been studied extensively in the past. While reviewing comprehensively
this large body of work is beyond the scope of this paper, we would 
like to note that a theoretically calculated formula from \citet{hm79}
and \citet{bh83} has been most commonly used in the past
\citep[see][for a recent review]{cs04}. In this work, following
our main ideology of using the most empirically determined quantities,
we adopt an observationally determined value for $R_d$ from
\citet{wthk08}. We also assume that the dust-to-gas ratio scales
linearly with gas metallicity $Z$ (in solar units) and that the gas is
clustered on scales 
unresolved in our simulations, so that the effective $\H2$ formation
rate is higher in proportion to the gas clumping factor,
$C_\rho\equiv\langle\rho^2\rangle/\langle\rho\rangle^2$,
\begin{equation}
  R_d = 3.5\times10^{-17} Z C_\rho \dim{cm}^3/\dim{s}.
  \label{eq:rd}
\end{equation}

The clumping factor $C_\rho$ can be viewed as a parameter of our
model. However, there is also some theoretical arguments for its
plausible range of values. Numerical simulations of turbulent molecular clouds typically
find a lognormal density distribution inside the clouds with the width
that depends on the average Mach number of the flows:
$\sigma_{\ln\rho}^2 \approx 1 - 2$ \citep{pjn97,osg01,mo07}. Since
for a lognormal distribution, 
\[
C_\rho = e^{\displaystyle\sigma_{\ln\rho}^2},
\]
most likely values for the clumping factor are in the range
$C_\rho\sim 3-10$. Empirical evidence also indicates that gas in
molecular clouds is highly clumped with the ratio of typical molecular
gas density to the mean cloud density of $\sim 30-100$ \citep[see \S
3.1.2 in][and references therein]{mo07}.

\begin{figure}[t]
\plotone{\figname{cdSobTrue.ps}{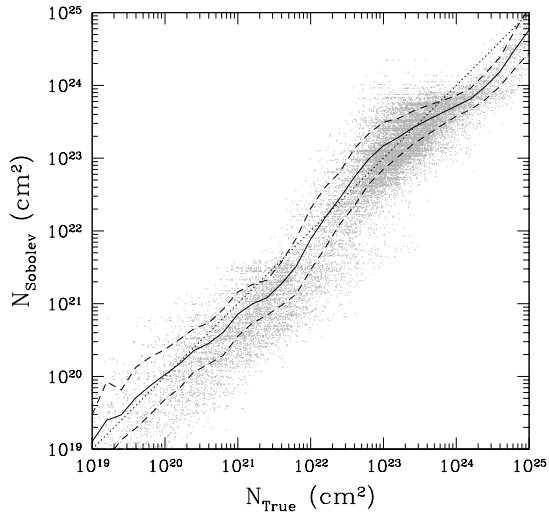}}
\caption{Comparison of the total hydrogen column density from the
  Sobolev-like approximation and the true column density as integrated
  along random lines of sight through the simulation box. Gray
  points show individual lines of sight; black solid line show the
  average $N_{\rm Sobolev}$ for a given $N_{\rm True}$, while black
  dashed lines give the rms scatter. The thin black dotted line is
  a diagonal of the plot.}
\label{fig:sob}
\end{figure}

Finally, in order to specify the column densities inside the molecular
clouds, we use a Sobolev-like approximation:
\begin{equation}
  N_i \approx n_i L_{\rm Sob},
  \label{eq:sob}
\end{equation}
where
\[
L_{\rm Sob} \equiv \frac{\rho}{\left|\nabla\rho\right|}.
\]
We have verified that equation (\ref{eq:sob}) provides an essentially
unbiased estimate for the true column density (obtained by integrating along
random lines of sight) within the range $3\times10^{20}\dim{cm}^2 <
N_\HI+2N_\H2 < 3\times10^{23}\dim{cm}^2$ with a scatter of about a
factor of 2, as illustrated in Figure \ref{fig:sob}.

In the gas dynamics solver and in the calculation of the cooling
function, we use 
the thermodynamically correct relations between the gas internal
energy, pressure, and temperature, {\it without assuming a polytropic
  approximation\/}, which becomes invalid for mostly molecular gas
(Turk, Gnedin, \& Abel, in preparation).

This model is implemented in the ART code. Because individual
molecular clouds are not resolved in our simulations, the specific
numerical implementation suffers from numerical artifacts due to
finite spatial resolution. We discuss these artifacts and our approach
for their mitigation in the Appendix.

\begin{figure}
\plotone{\figname{rf.ps}{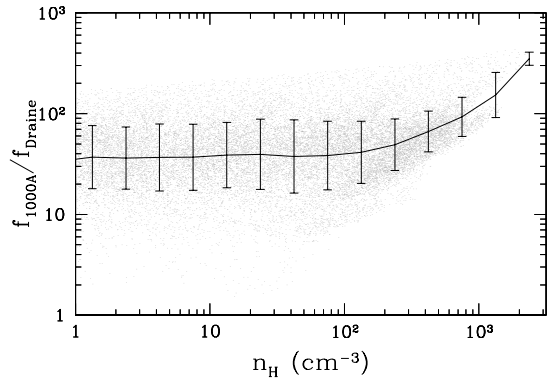}}
\caption{\label{fig:rf} The intensity of the interstellar radiation
  field at $1000\AA$ as a function of gas density in the parent
  simulation FNEC-RT, in units of the \citet{h2:d78} field
  ($10^6\dim{photons}/\dim{cm}^2/\dim{s}/\dim{sterrad}/\dim{eV}$).
  Since the parent simulation does not include dust shielding,
  radiation at $1000\AA$ in that simulation was computed in the
  optically thin regime.}
\end{figure}

\subsection{Star Formation}

We explore three simple star formation recipes, based on the local
density of molecular hydrogen, in order to test the dependence of
global KS relations on the underlying assumptions. We can write the
star formation rate in a molecular cloud in terms of an efficiency
factor $\epsilon$, equivalent to the inverse of the (molecular) gas
consumption timescale, as 
\beq
\dot{\rho}_\star = \epsilon \rho_{\H2},
\eeq
where $\dot{\rho}_\star$ is the star formation rate per unit volume, and
$\rho_{\H2}$ is the molecular hydrogen density.
We can then express $\epsilon$ in terms of a dimensionless
effective efficiency $\varepsilon_{\rm ff}$ per local free-fall time
of the gas, so that 
\beq
\dot{\rho}_\star = \frac{\varepsilon_{\rm ff}}{\tau_{\rm sf}} \rho_{\H2}, 
\eeq
where $\tau_{\rm sf} = \tau_{\rm ff} = \sqrt{3\pi/32G\rho}$ for a
uniform sphere. Since stars form out of regions of molecular complexes
with sizes not necessarily resolved in our simulations, the choice of
a relevant star formation timescale relies on assumption about the
structure of the gas in sub-grid scales. We have tried two different
assumptions about $\tau_{\rm sf}$: 
\begin{description}
\item[SF1] the constant time scale $\tau_{\rm sf} = \tau_{\rm
  ff}(100\dim{cm}^{-3})$, assuming all $\H2$ in a grid cell resides in
  molecular clouds of average density $\approx 100\dim{cm}^{-3}$;
\item[SF2] the time scale, which depends on the cell's gas density
  with an upper limit $\tau_{\rm sf} = \min[\tau_{\rm
  ff}(100\dim{cm}^{-3}),\tau_{\rm ff}(\rho_{g,cell})]$.
\item[SF3] the time scale, which simply depends on the density of the
  cell $\tau_{\rm sf} = \tau_{\rm ff}(\rho_{g,cell})$ (this recipe is
  similar to the one used by \citet{robertson_kravtsov08} in their
  models of isolated galaxies).

\end{description}

We adopt $\epsilon_{\rm ff} = 0.01$ \citep{krum} in all of the
calculations presented below; we also use a molecular fraction
threshold for computational efficiency: star
formation is allowed only in cells which have a molecular fraction
higher than $f_{\H2}=0.1$. The latter criterion has a negligible
impact on the total star formation rate in a galaxy, but is highly
beneficial from a computational point of view, avoiding formation of a
large number of very small stellar particles in low density regions.

\section{Results}\label{res}

In this section we present results illustrating the dependence
of the relative atomic and molecular gas abundances on metallicity, gas
clumpiness, 
and UV flux. We also explore the correspondent dependencies of
the global Kennicutt-Schmidt relations with different local star 
formation prescriptions described in the previous section. 

\begin{deluxetable}{lcccc}
\tablecaption{Simulations\label{table1}}
\tablewidth{0pt}
\tablehead{
\colhead{Simulation} & \colhead{Metallicity} & \colhead{Clumping} & 
\colhead{UV flux}& \colhead{SF}\\
\colhead{} & \colhead{$Z/Z_\odot$} & \colhead{Factor $C_\rho$} & 
\colhead{}& \colhead{Recipe}
}
\startdata\\
A & $1$ & $10$  & $1$ & SF2\\
B1 & $0.3$  & $10$ & $1$ & SF2\\
B2 & $0.1$ & $10$ & $1$ & SF2\\
C1 & $1$ & $1$ & $1$& SF2\\
C2 & $1$ & $100$ & $1$& SF2\\
D & $1$ & $10$  & $10$ & SF2\\
E & $1$ & $10$  & $1$ & SF1\\
\enddata
\end{deluxetable}

Using the $z=4$ output of the parent FNEC-RT simulation of
\citet{tkg08} (see \S~\ref{dsims}) as initial condition, we run a
simulation with the model for molecular hydrogen and star formation
for $100\dim{Myrs}$. The baseline simulation FNEC-RT does not
incorporate shielding of molecular hydrogen, and thus no molecular
clouds form in that simulation. The chosen interval of time of
$100\dim{Myr}$ is considerably longer than the characteristic lifetime
of molecular clouds \citep[$\sim 10^7$~yrs,][]{bshu80, blitzetal07}
and is therefore sufficiently long for our purposes. We have also
verified that our results have converged: various distribution
presented below are virtually indistinguishable if we chose a
$50\dim{Myr}$ time interval instead.

The baseline simulation FNEC-RT also includes effects of both energy
and metal feedback to the interstellar medium, to keep the tests clean 
both forms of feedback are turned off and the interstellar medium
metallicity is kept at a constant value after our test simulation is
started. 

In our fiducial simulation (A) the metallicity is kept constant at
solar value, we use a clumping factor $C_p=10$, and the star formation
recipe used is SF2. The UV flux is taken directly from the parent
simulation FNEC-RT and is shown in Figure \ref{fig:rf}. Our model
galaxy has a significantly higher star formation rate and is smaller
at $z=4$ galaxy than the Milky Way today, so the interstellar
radiation field is substantially higher than the canonical
\citet{h2:d78} value adopted for the Milky Way ISM at the present
time, but is consistent with the estimates of the interstellar UV
field in high redshift galaxies \citep{cppp09}.

In each of the simulations used in our parameter study, we vary a
single parameter away from its fiducial simulation value. Table
\ref{table1} describes the values of these free parameters in each of
the simulations discussed in this section. 

\begin{figure}
\plotone{\figname{gal6.ps}{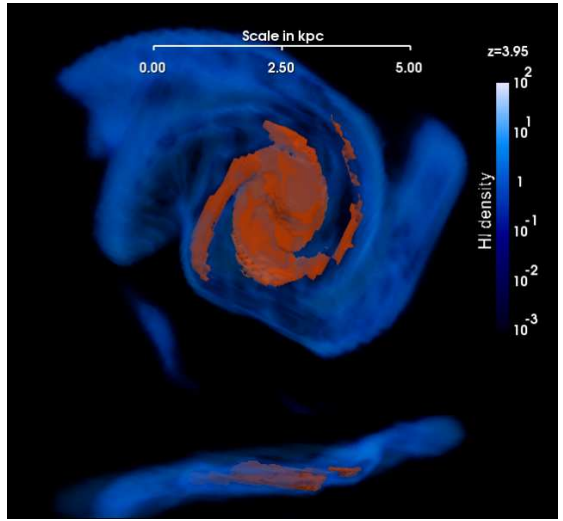}}
\caption{\label{fig:image} A face- and edge-on views of the largest
  galaxy in our fiducial simulation A. Blue translucent color
  represents volume rendering of atomic hydrogen density, while orange
  color shows the location of the molecular gas (as the isosurface of
  $f_\H2=0.5$ value). This figure is best viewed in color.}
\end{figure}

A visual representation of our fiducial simulation A is shown in
Figure \ref{fig:image}. The molecular gas traces spiral arms well,
which are less pronounced in the atomic gas. The molecular disk is
both smaller and thinner than the atomic disk. However, since our
spatial resolution is only $50\dim{pc}$, we do not resolve individual
molecular clouds. Instead, the orange colored surface in the image
shows the boundary of the mostly molecular ($f_\H2>0.5$) gas.

\subsection{Gas phases and atomic-to-molecular transition 
in the ISM of model galaxies}

\begin{figure}
\plotone{\figname{nT_rho_Z_bin_C6.1.eps}{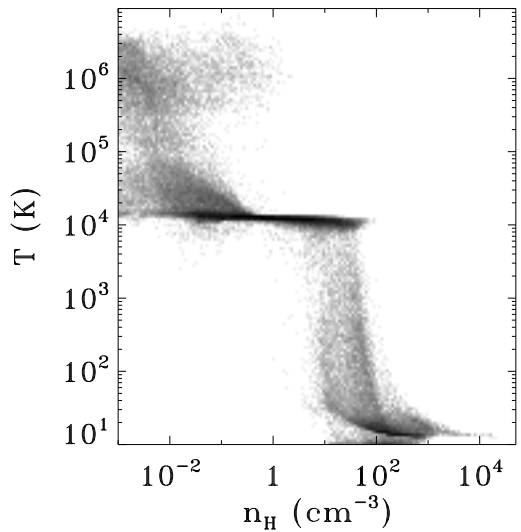}}
\caption{\label{T_rho_C6} Temperature plotted against the total gas
  number volume density for all maximum refinement level (level 9)
  cells at $z=4$, for simulation A.} 
\end{figure}

Figure~\ref{T_rho_C6} shows the temperature in degrees Kelvin plotted
against the total volume number density of neutral hydrogen for our
fiducial simulation (A). Points in this and subsequent plots in this
subsection show cells at the highest resolution level (level 9; the
physical size of 52~pc at $z=4$), which cover the large
fraction of the disks of galaxies forming in the high-resolution
Lagrangian region of the simulation. This plot clearly shows a well developed
multi-phase structure of the ISM in these disks. In particular, three 
different gas phases are evident: (i) the hot, ionized, low-density
gas in the upper left part of the diagram; (ii) the warm neutral
medium around $T\sim 10^4\dim{K}$ and $n_{\rm H}\sim
0.1-10\dim{cm}^{-3}$ and (iii) the cold neutral medium with
$T\sim10-100\dim{K}$ at $n_{\rm H}>100\dim{cm}^{-3}$. The transition from
the warm neutral to cold neutral phase occurs over a narrow range of
gas densities around a few tens $\dim{cm}^{-3}$, in good agreement
with the models of the Milky Way interstellar medium (which has
similar metallicity to our model) of \cite{Wolf03}.

\begin{figure*}
\plotone{\figname{nfH2_Z_bin_C6.1.eps}{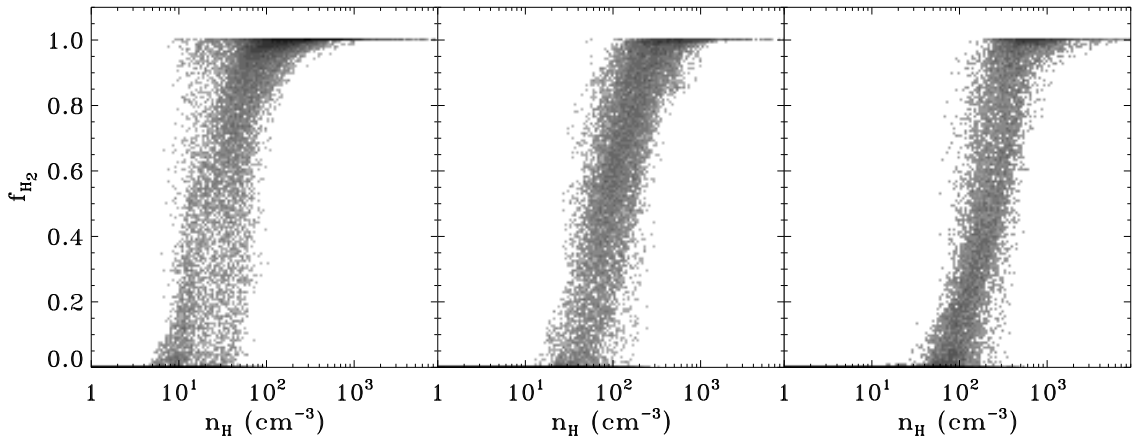}}
\caption{\label{nfH2_Z_C6} Molecular hydrogen fraction in level 9
  cells, plotted against the total neutral gas number volume
  density. The grayscale corresponds to the density of binned points
  on the plane. Left panel: simulation A (solar metallicity); middle
  panel: simulation B1 (metallicity $0.3$ solar); right panel:
  simulation B2 (metallicity $0.1$ solar).} 
\end{figure*}

Although not apparent in Figure~\ref{T_rho_C6}, the cold neutral medium 
undergoes a transition from atomic to fully molecular phase at the
gas density of $n_{\rm H}\approx 100\dim{cm}^{-3}$. Figure \ref{nfH2_Z_C6}
shows the molecular fraction, 
$f_\H2=2n_\H2/\left(n_\HI+2n_\H2\right)$, as a function of the total neutral
hydrogen volume number density. The three panels correspond to
simulations with decreasing metallicity (left panel: simulation A,
$Z=Z_{\odot}$; middle panel: simulation B1, $Z=0.3Z_{\odot}$;
right panel: simulation B2, $Z=0.1Z_{\odot}$).  A sharp transition from fully atomic to fully molecular
gas occurs at high gas densities, with the characteristic density of the 
transition increasing with decreasing metallicity. The gas density
at which the molecular fraction is 50\% scales with metallicity approximately as
\beq
n_{\rm t} \simeq 30 \left( \frac{Z}{Z_\odot} \right)^{-1} \,\, {\rm cm^{-3}}.
\eeq
This trend is not surprising, as the
atomic to molecular transition is facilitated by the dust opacity:
when dust opacity is sufficiently high to shield the gas from UV radiation,
the transition to molecular medium can occur (self-shielding only
becomes dominant at higher molecular fractions). As the metallicity
increases, the amount of dust available to shield the gas from UV also
increases, and the atomic-to-molecular transition can occur at lower
densities. 

\begin{figure*}
\plotone{\figname{f6by.eps}{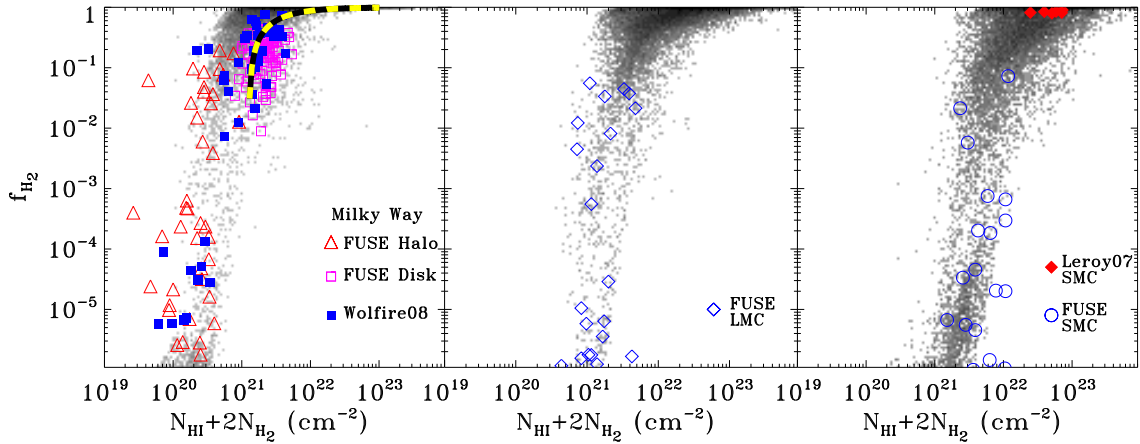}}
\caption{\label{nfH2_Z_col_C6} Molecular hydrogen fraction in level 9
 cells, plotted against the total neutral gas column
 density. The grayscale corresponds to the density of binned points on
 the plane. Left panel: simulation A (solar metallicity); middle
 panel: simulation B1 (metallicity $0.3$ solar); right panel:
 simulation B2 (metallicity $0.1$ solar). Different symbols shows
 observational data, as follows: open circles: FUSE SMC measurements
 (compiled by \citet{h2:bts03} and \citet{h2:gstd06}); filled
 diamonds: SMC measurements from \citet{lbsm07}; open diamonds: FUSE LMC
 measurements; open triangles: FUSE MW Halo measurements; open squares: FUSE MW
 disk measurements; filled squares: MW measurements from
 \cite{wthk08}. The black/yellow striped line in the left panel shows the
 location of points with $\Sigma_\HI=10\msunpc2$.}
\end{figure*}

To compare these results with observational measurements in
Figure~\ref{nfH2_Z_col_C6} we plot the molecular fraction against the
total neutral hydrogen {\em column} density, using a logarithmic
vertical scale, for simulations A, B1, B2. We overplot the FUSE
measurements of the molecular fraction in the MW halo and disk (left
panel), in the LMC (middle panel), and in the SMC (right panel)
compiled by \citet{h2:bts03} and \citet{h2:gstd06}. In the left panel
we also plot data for the Milky Way from \cite{wthk08}, and on the
right panel we show the data for SMC from \citet{lbsm07}. The
metallicities of the ISM in these galaxies approximately correspond to
the metallicities adopted in the models against which we compare
them. In all cases, the column density at which the molecular
fractions begin to increase rapidly and the sharpness of the
transition are well reproduced in our simulations. For the solar
metallicity case, this transition corresponds to the maximum surface
density of atomic hydrogen of about $10\msunpc2$, in
concordance with the measurements of molecular and atomic gas surface
densities in nearby galaxies
\citep{wong_blitz02,blitz_rosolowsky06,bigiel_etal08}. The observed
trend of the atomic-to-molecular transition 
column density to decrease with increasing metallicity is also
quantitatively reproduced in our calculations.

\begin{figure}
\plotone{\figname{fhicd2.ps}{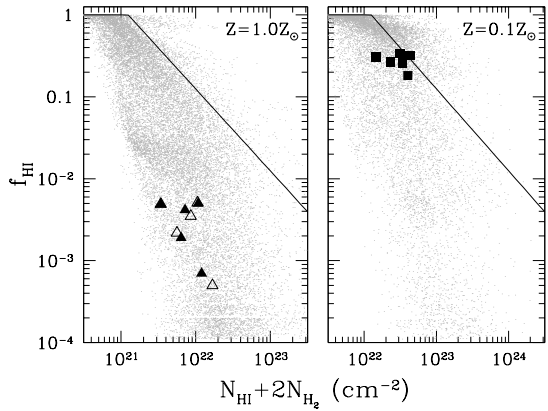}}
\caption{\label{fig:fhicd} Atomic hydrogen hydrogen fraction in level 9
  cells, plotted against the total gas number density (a complementary
  figure to Fig.\ \ref{nfH2_Z_col_C6}). Two panels show solar and 10\%
  solar metallicity cases (we have found no observational data for
  the LMC case, so we do not show 30\% solar metallicity case). Open
  and filled triangle for the solar metallicity case show measurements
  from \citet{GL05}, while filled squares in the 10\% solar
  metallicity panel show \citet{lbsm07} data. Solid lines in both
  panels show constraints $\Sigma_\HI=10\msunpc2$ and
  $100\msunpc2$ respectively.}
\end{figure}

Figure\ \ref{nfH2_Z_col_C6} does not illustrate the relationship between
the atomic and molecular gas inside mostly molecular gas, when
$f_\H2\approx1$. In order to test our model in that regime, we show in 
Figure \ref{fig:fhicd} a complement of Figure\ \ref{nfH2_Z_col_C6},
the atomic hydrogen fraction as a function of the total gas column
density. While the measurements of atomic hydrogen fraction in the
mostly molecular gas are not numerous \citep{GL05,lbsm07}, they
provide a useful constraint for our model (see also the Appendix) in
the regime where the gas is mostly molecular ($f_\HI\ll1$).

\begin{figure*}
\plotone{\figname{nfH2_C_bin_C6.1.eps}{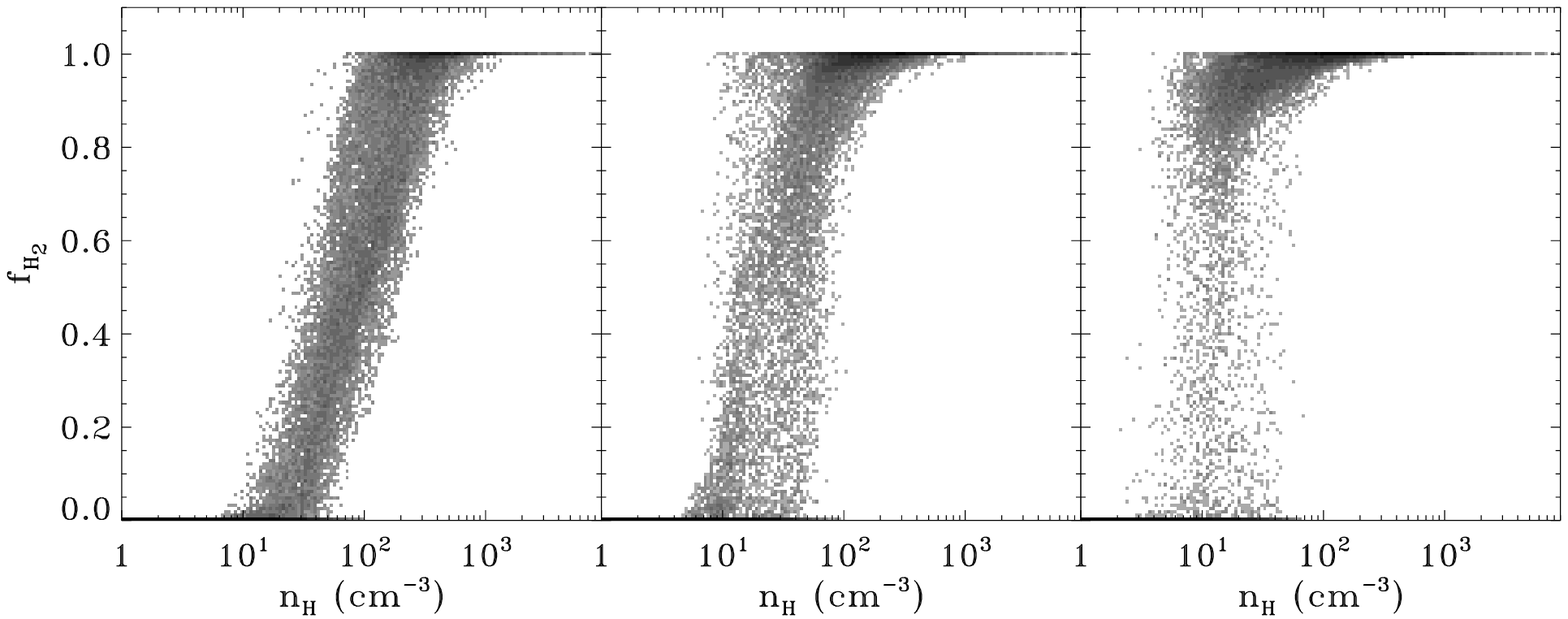}}
\caption{\label{nfH2_C_C6} Molecular hydrogen fraction in level 9
  cells, plotted against the total gas number density. Panels from the
  left to right correspond to clumping factor $C_\rho=1$
  (simulation C1), $C_\rho=10$ (simulation A), and $C_\rho=100$
  (simulation C2) respectively. The grayscale corresponds to the
  density of binned points on the plane.}
\end{figure*}

Figure~\ref{nfH2_C_C6} shows the dependence of the transition from
atomic to molecular phase on the adopted clumping factor. Although 
the dependence of the transition on the clumping
factor\footnote{And hence on the unresolved structure of the gas within our
resolution elements.} is weaker than on metallicity, it is still appreciable. 
Increasing the clumping factor steepens the transition and shifts it to 
lower number densities. Comparison with observations in Figure~\ref{nfH2_Z_col_C6}
indicates that the clumping factor of $C_{\rho}=10$ is a good fiducial
value. This value is also within the range of clumping factors expected
for log-normal PDFs in the theoretical models of molecular clouds (see \S~\ref{sec:H2model}).

\begin{figure*}
\plotone{\figname{nfH2_fuv_bin_C6.1.eps}{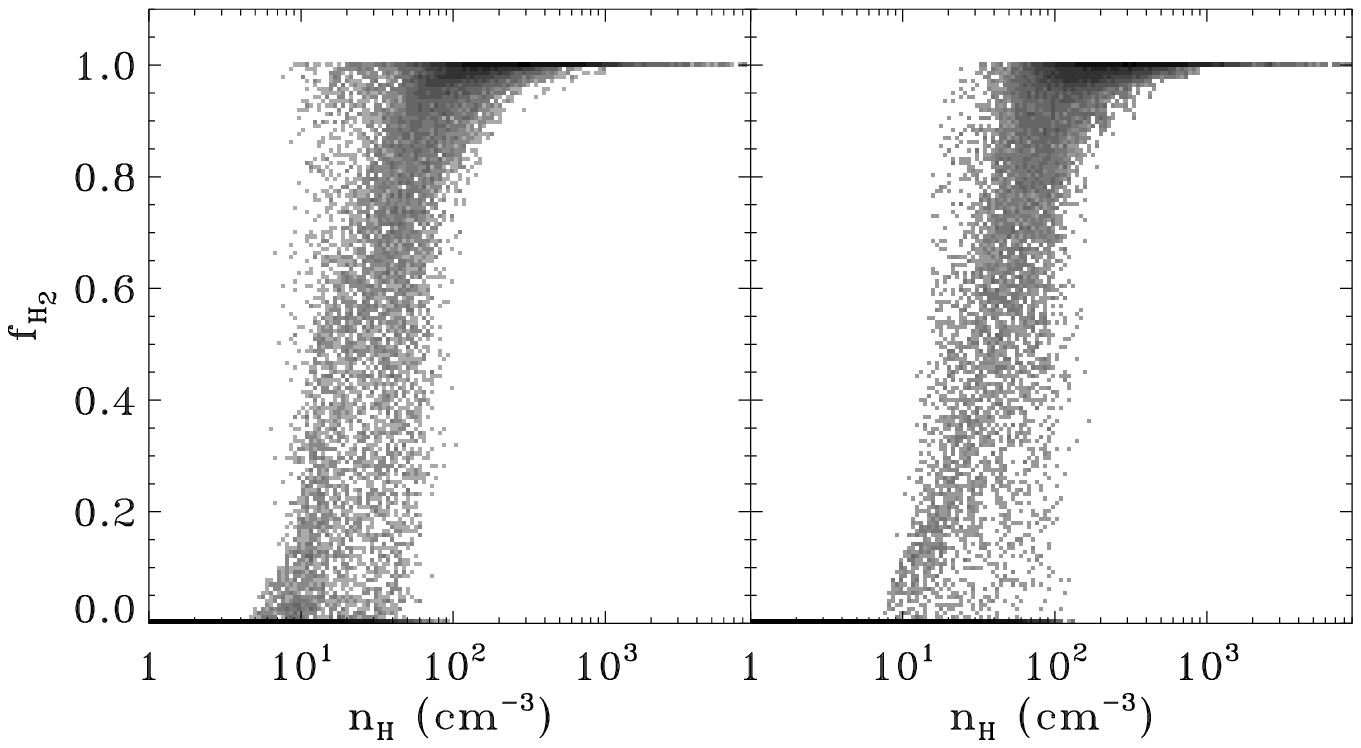}}
\caption{\label{nfH2_fuv_C6} Molecular hydrogen fraction in level 9
  cells, plotted against the volume number density of the total
  gas. The right panel corresponds to a simulation with 10 times the
  UV flux (simulation D) of that on the left (simulation A). The
  grayscale corresponds to the density of binned points on the plane.} 
\end{figure*}

\begin{figure*}
\plotone{\figname{nfH2_tsf_bin_C6.1.eps}{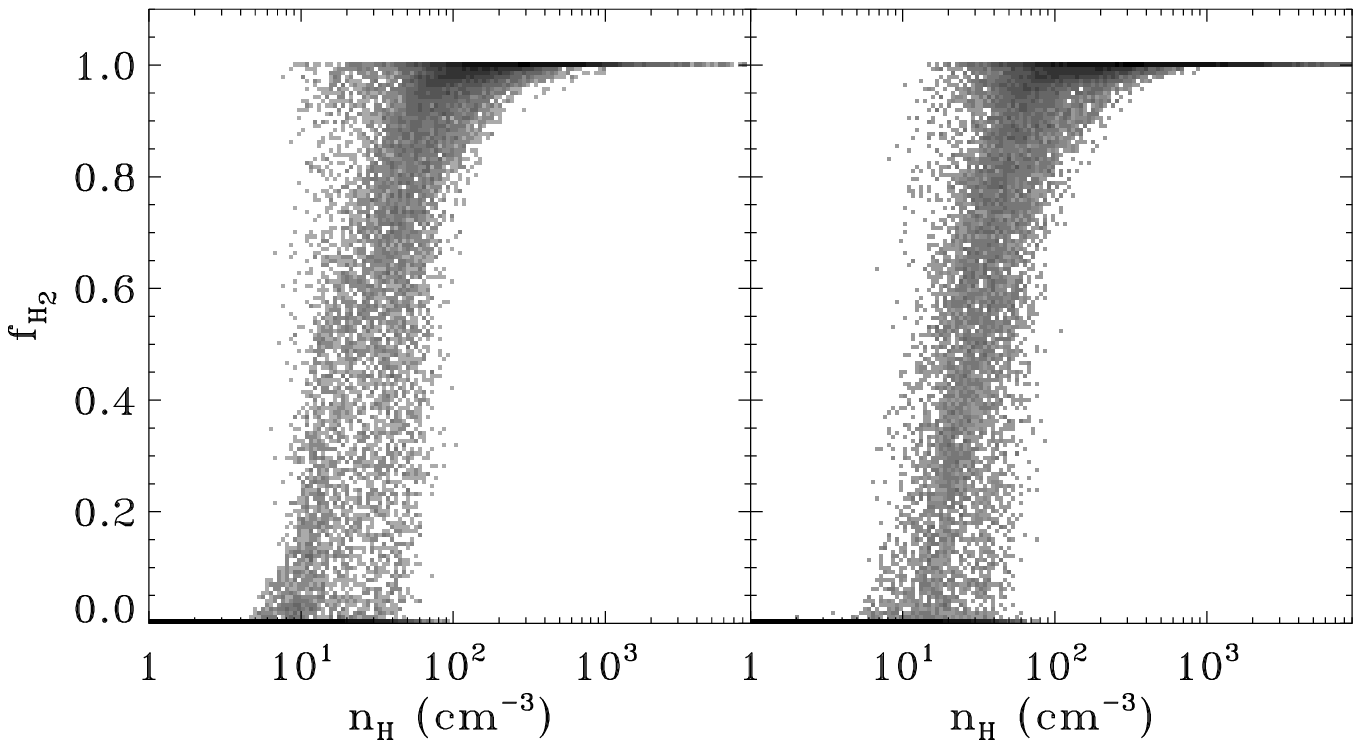}}
\caption{\label{nfH2_tsf_C6} Molecular hydrogen fraction in level 9
 cells, plotted against the total gas number density. Left panel:
 simulation A (star formation recipe SF2); right panel: simulation E
 (star formation recipe SF1). The grayscale corresponds to the density
 of binned points on the plane. } 
\end{figure*}

In Figure~\ref{nfH2_fuv_C6} we examine the dependence of the transition from 
atomic to molecular gas on the UV flux. The right panel (simulation D)
represents a UV flux 10 times greater than the one on the left panel
(simulation A); that value is at the upper range of observational
estimates of the UV field in gamma-ray burst hosts
\citep{cppp09}. Although some differences can be seen, they are 
much smaller than in the previous figures. The metallicity of the gas,
rather than the amount of dissociating radiation, is the primary
factor determining the density of atomic to molecular transition. 
This result may appear counter-intuitive at first, but it is not surprising. Since the UV flux is high, in the optically thin regime the equilibrium fraction of molecular hydrogen is of the order of
$10^{-8}$. Thus, gas becomes mostly molecular only when shielding
optical depth is of the order of $-\ln(10^{-8}) \approx 16$. At this
high column density, a factor of 10 increase in the UV flux (and,
thus, a factor of 10 decrease in the equilibrium $\H2$ fraction in the
optically thin regime) can be fully compensated by the increase of the
optical depth (and, thus, the column density)  to 
$-\ln(10^{-9}) \approx 18$, which can be achieved, for example, by
only $\approx 15\%$ change in the gas density. 

This is not to say that the UV flux is not at all important. Recall
that the mean 
flux within our model ISM is already quite high (Figure~\ref{fig:rf}). 
Without this radiation the abundance of $\H2$ at smaller densities
would be quite high \citep{robertson_kravtsov08} and transition 
to fully molecular phase not as steep as indicated by observations
(Figure~\ref{nfH2_Z_col_C6}). The role of the interstellar UV flux 
therefore appears to be in controlling the amount of diffuse $\H2$
and maintaining the thermodynamic balance of the cold neutral 
medium at lower densities, while the transition to the fully molecular
phase is controlled primarily by metallicity and clumpiness of the gas.

For completeness, Figure~\ref{nfH2_tsf_C6} shows dependence of the 
atomic to molecular phase transition on the adopted star formation recipe. In particular, 
we compare our fiducial simulation (A) with a star formation
timescale which scales with the inverse square root of the gas
density (SF2) and simulation E with a constant star formation
timescale (SF1). The figure shows that there is no discernible
differences between these two recipes (we have also checked that there
is no difference when SF3 recipe is adopted). The details of the star
formation recipe adopted thus do not affect the transition of atomic to 
molecular phase, at least within the range of recipes and parameters
we considered.

\begin{figure*}
\plotone{\figname{C6.1SFLAW_l3_Z_e1.eps}{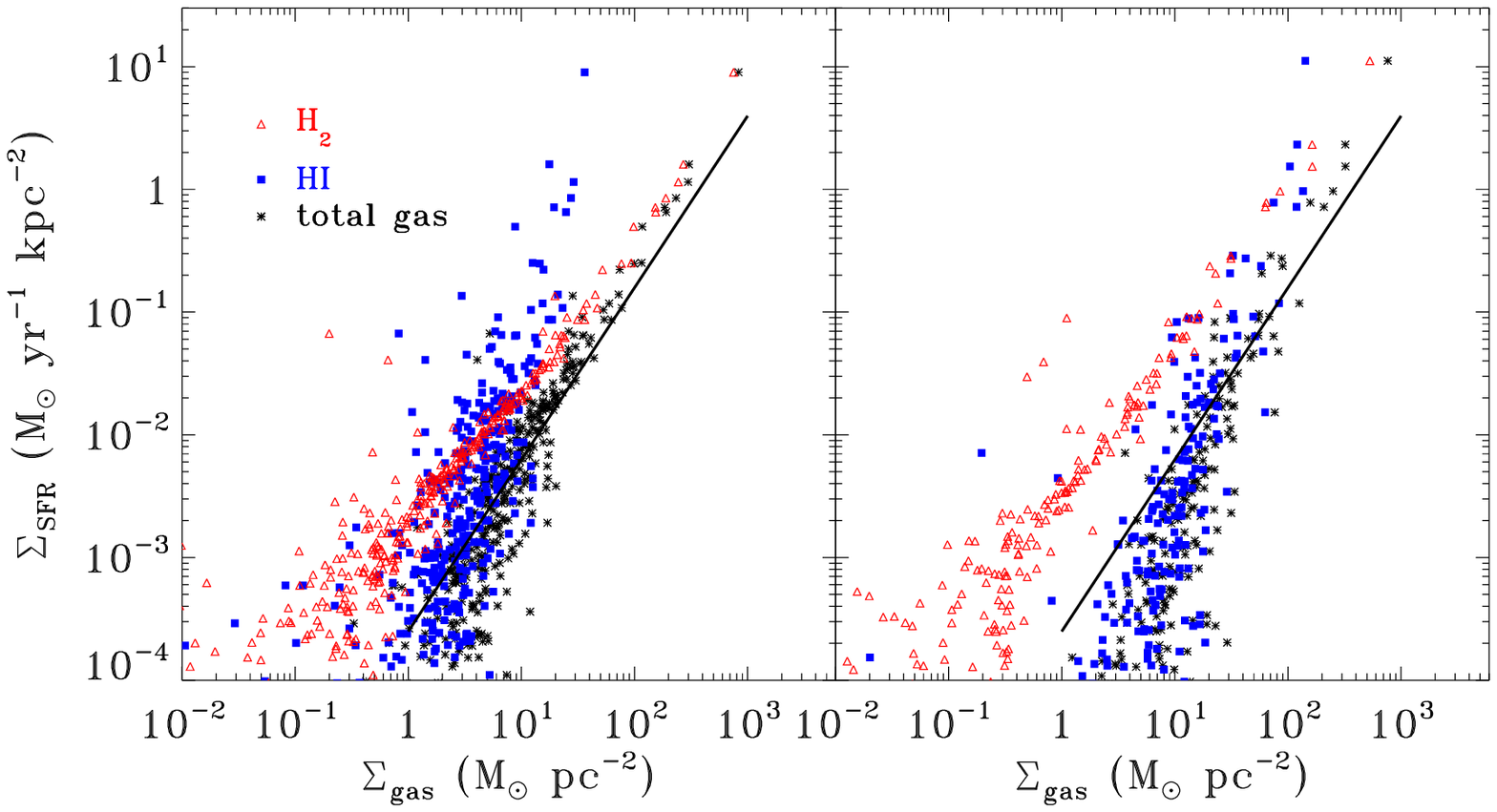}}
\caption{\label{NSFLAW_l3_Z_C6} 
Sensitivity of the $\Sigma_{\rm SFR} - \Sigma_{\rm gas}$ relation to
metallicity. Quantities are averaged over level 3 cells
(corresponding, at $z=4$, to a proper scale of $3.2$ kpc). Star
formation rates are averaged over 30 Myrs. Each point corresponds to a
different level 3 cell in the simulation at $z=4$. Left panel:
simulation A (solar metallicity); right panel: simulation B2
(metallicity 0.1 solar). Red triangles: molecular gas; blue squares:
atomic gas; black stars: total gas (no He). Solid line: \cite{Ken98}
law (slope equal to $1.4$, see Eq. \ref{KenEq}).}
\end{figure*}

\subsection{The Kennicutt-Schmidt relation}
\label{sec:ks}

\begin{figure*}
\plotone{\figname{C6.1SFLAW_l3_C_e1.eps}{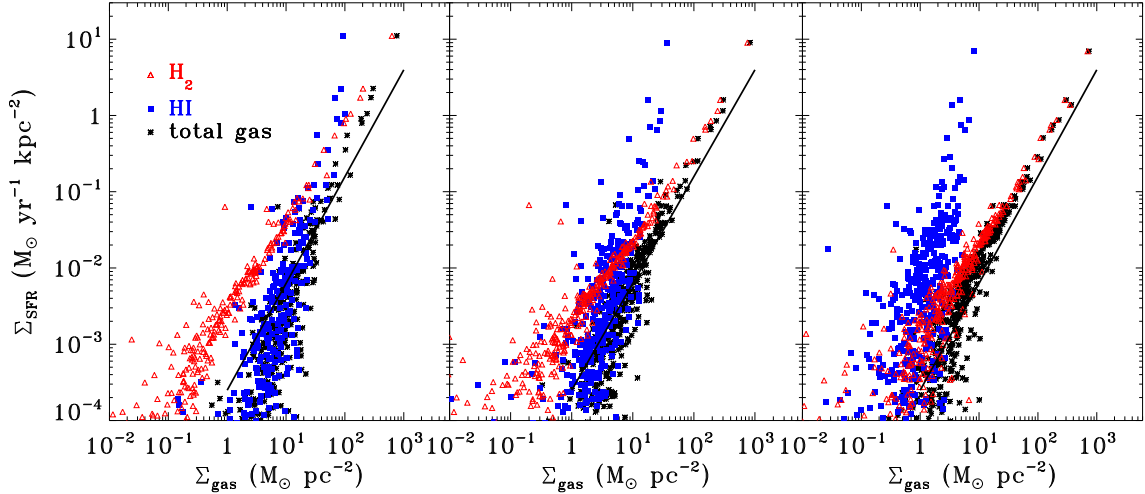}}
\caption{\label{NSFLAW_l3_C_C6} Sensitivity of the $\Sigma_{\rm SFR} -
 \Sigma_{\rm gas}$ relation  to clumping factor. Points and lines as
 in Fig.\ \ref{NSFLAW_l3_Z_C6}.  Left panel: $C_p=1$ (simulation C1);
 middle panel: $C_\rho=10$ (simulation A); right panel: $C_p=100$
 (simulation C2).  }
\end{figure*}

\begin{figure*}
\plotone{\figname{C6.1SFLAW_l3_fuv_e1.eps}{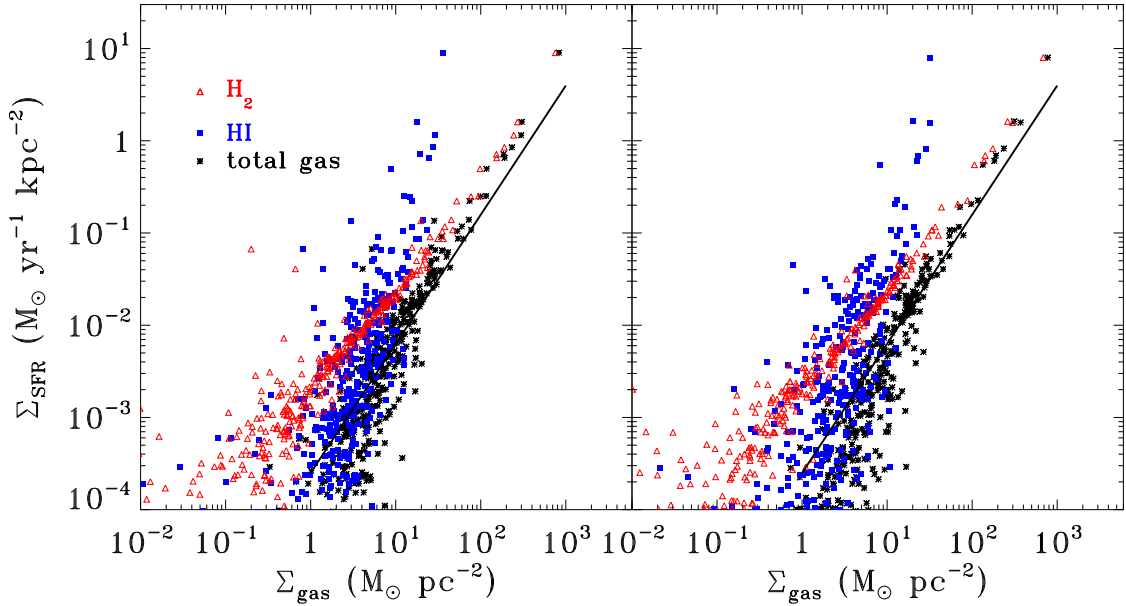}}
\caption{\label{NSFLAW_l3_fuv_C6} Sensitivity of the $\Sigma_{\rm SFR}
 - \Sigma_{\rm gas}$ relation  to the UV flux. Points and lines as in
 Fig.\ \ref{NSFLAW_l3_Z_C6}. The right panel corresponds to a
 simulation  with 10 times the UV flux (simulation D) than that on the
 left (simulation A).} 
\end{figure*}

\begin{figure*}
\plotone{\figname{C6.1SFLAW_l3_sfr_e1.eps}{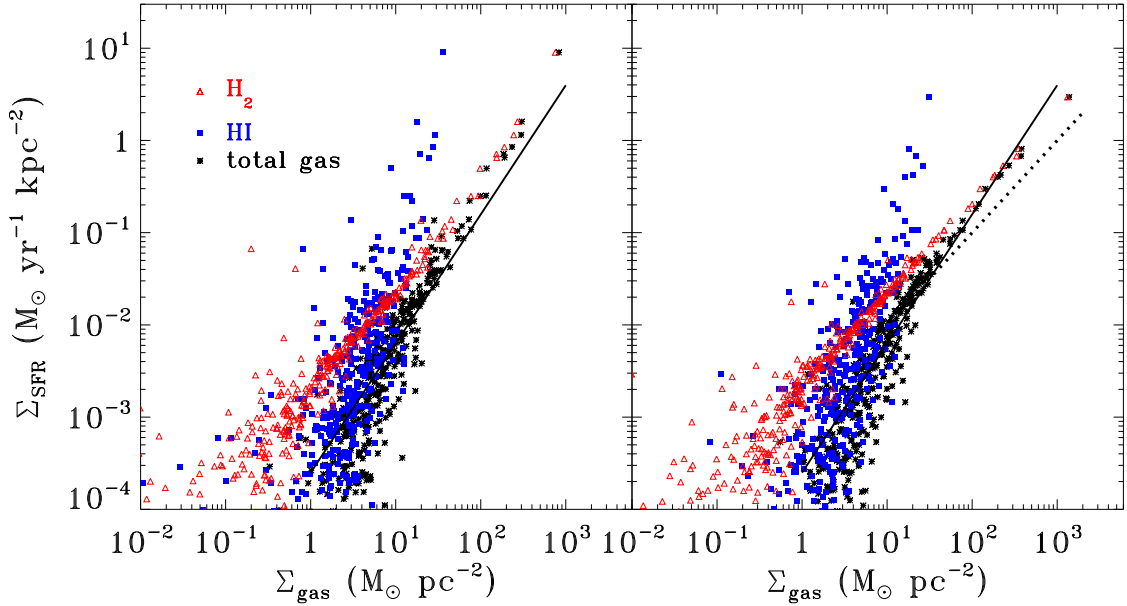}}
\caption{\label{NSFLAW_l3_sfr_C6} 
Sensitivity of the $\Sigma_{\rm SFR} - \Sigma_{\rm gas}$ relation to the star 
formation recipe. Points and lines as in Fig.\ \ref{NSFLAW_l3_Z_C6}. 
Left panel: star formation recipe SF2 (simulation A); right panel: star 
formation recipe SF1 (simulation E). To illustrate the difference in
the slope better, we also show a KS relation with the slope of 1 for
$\Sigma>30\msunpc2$. with a dotted line.} 
\end{figure*}

In this section we examine the large-scale relation between star
formation and gas surface densities (the Kennicutt-Schmidt relation)
in the simulated galaxies with the molecular gas model and star
formation prescriptions discussed in the previous sections.
Figures \ref{NSFLAW_l3_Z_C6} - \ref{NSFLAW_l3_sfr_C6} show this
relation for different choices of the model parameters and star
formation prescriptions. Similarly to observational estimates, 
the star formation rate in this figure was
averaged over 30~Myr, the time interval comparable to the
characteristic life time of molecular clouds and massive stars. Each
point in the figures corresponds to a level 3 cell in the simulation
at $z=4$; the physical cell size is 3.3~kpc and thus both the star
formation and gas surface densities have been averaged on this
scale. In addition to the total gas surface density, the figures also
show the dependence of star formation on the surface density of atomic
and molecular gas separately.  The solid line in each panel
corresponds to the best fit to the empirical correlation estimated for
nearby massive and starburst galaxies \cite{Ken98}: 
\beq\label{KenEq}
\Sigma_{\rm SFR} = 2.5 \times 10^{-4} \left(\frac{\Sigma_{\rm
gas}}{\rm 1\msunpc2}\right)^{1.4} {\rm \, M_\odot \,
yr^{-1} kpc^{-2}}\,.  
\eeq 

Before we discuss these figures further, it is worth explaining
certain features of these plots which may look peculiar to the
reader. When one compares the KS relations with different $\H2$ and SF
model parameters, $\Sigma_{\rm SFR}$ sometimes does not change, while
$\Sigma_\HI$ and $\Sigma_\H2$ change dramatically. This seemingly
strange behavior is particularly apparent at high surface
densities. The reason is that star formation rate is averaged over
$30\dim{Myr}$, while the fraction of $\H2$ and $\HI$ can change on a
shorter time scale. Thus, if a region of high surface density becomes
fully molecular and forms a population of stars, which then dissociate
much of the remaining $\H2$, $\Sigma_{\rm SFR}$ will reflect the fully
molecular $\Sigma_{\rm gas}$ and not the current instantaneous
$\Sigma_\H2$. The latter reflects the speed with which the molecular
gas is able to regenerate after being dissociated by young stars. This
speed is controlled by the metallicity and clumpiness of the medium
and thus the instantaneous $\Sigma_\H2$ will reflect these
dependencies, while $\Sigma_{\rm SFR}$ will not.

Figure \ref{NSFLAW_l3_Z_C6} shows that the KS relation is quite
sensitive to the metallicity of the gas. The left panel corresponds to
our fiducial simulation (A) with solar metallicity and the right panel
to simulation B2 with metallicity 0.1 solar. While the star formation
rate is tightly correlated with the total and $\H2$ surface
densities, the correlation with the $\HI$ density exhibits considerable
scatter. In addition, the $\HI$ surface density saturates at a relatively
low surface density, which depends on metallicity (at $\approx
30\msunpc2$ for $Z=Z_{\odot}$ and $\approx 100\msunpc2$ for
$Z=0.1Z_{\odot}$). Such saturation is 
consistent with observations \citep[see, e.g.,][]{ken07, bigiel_etal08,
lb08}. The surface density at which $\HI$ saturates and its metallicity
dependence are of course related to the volume density of the
transition from the atomic to molecular phase (see
Fig.~\ref{nfH2_Z_col_C6}). Given that the latter depends on the choice
of the model parameters, the saturation in the KS relation can be used
as an additional observational constraint.

The relationship between the star formation rate and the molecular gas
surface density (``molecular KS relation'') is less affected by the
gas metallicity. Indeed, at a given surface density, the star
formation rate is about 3 times higher in a low metallicity gas. This
simply reflects the fact that, with our assumptions of a linear
relation between the dust-to-gas ratio and gas metallicity, gas at 0.1
solar metallicity forms stars at typical densities that are 10 times
higher than in a solar metallicity gas, resulting in about a factor of
3 higher specific star formation rate for our fiducial star formation
recipe SF2; respectively, there would be almost no change in the
molecular KS relation for the SF1 recipe.

Figure~\ref{NSFLAW_l3_C_C6} compares the KS relations for the
calculations with different clumping factors, $C_\rho$. Our fiducial
simulation A ($C_\rho=10$) is shown in the middle panel, while a
simulation with smaller and greater clumping factors (simulation C1
with $C_\rho=1$ and simulation C2 with $C_\rho=100$) are shown in the
left and right panels respectively.  Qualitatively, the results are
similar to those in Figure~\ref{NSFLAW_l3_Z_C6}.  However, the figure
shows significant sensitivity of the $\HI$ saturation surface density to
the clumping factor. Again, this is related to the corresponding
dependence of the local volume density of the transition shown in
Figure~\ref{nfH2_C_C6}. The $\HI$ saturation surface density observed in
nearby galaxies can place constraints on the value of the clumping
factors. For example, the gas in M51 has metallicity of $Z\approx
2Z_{\odot}$ and its KS relation exhibits saturation at
$\Sigma_\HI\approx 30\msunpc2$.  In a larger
sample of 
both massive spirals and dwarf galaxies (with typically $Z<Z_{\odot}$)
reported by \citet{lb08}, the $\HI$ saturation occurs at
$\Sigma_\HI\approx 10-20\msunpc2$.  Comparing
these 
numbers to the KS relations in Figure~\ref{NSFLAW_l3_C_C6}, indicates
that the clumping factor required to reproduce these observations
(especially at lower than solar metallicities) is quite high,
$C_{\rho}\gtrsim 10$. This is, in fact, consistent with observational
estimates of the clumping within molecular clouds which indicate 
that typical ratios of the local to the mean density of a cloud are
$\sim 30-100$ \citep[see \S
3.1.2 in][and references therein]{mo07}.

 As Figure~\ref{nfH2_C_C6} demonstrates, the column density at which the $\HI$
to $\H2$ transition occurs decreases with increasing clumping factor;
as a result, the surface density at which $\HI$ saturates decreases with
increasing clumping factor. Consequently, the star formation rate
scaling with the $\H2$ gas approaches the scaling with the total gas
as the clumping factor increases, and become very similar for the
highest clumping factor we have used ($C_\rho=100$).

The sensitivity of the KS relation to the UV flux is shown in
Figure \ref{NSFLAW_l3_fuv_C6}. The right panel corresponds to a
simulation with 10 times the UV flux (simulation D) than that that of
our fiducial run on the left (simulation A). No appreciable changes
can be seen between the two panels. The dependence of the global
star formation on the UV flux is therefore much weaker than on 
the metallicity and clumpiness. As we have noted above, this
does not mean that the UV flux is not important at all. 
With the UV flux close to zero the abundance of $\H2$ at smaller densities
would be quite high \citep{robertson_kravtsov08} and transition 
to fully molecular phase not as steep as indicated by observations
(Figure~\ref{nfH2_Z_col_C6}). The role of the interstellar UV flux 
therefore appears to be in controlling the amount of diffuse $\H2$
and maintaining the thermodynamic balance of the cold neutral 
medium at lower densities. Beyond certain level, however, 
the dependence of the results on the UV flux saturates. 

Finally, Figure \ref{NSFLAW_l3_sfr_C6} shows the dependence of the 
scaling on the local star formation 
recipe used. The left panel corresponds to our fiducial simulation (A),
with star formation recipe SF2 (timescale scaling with density), 
while the right panel corresponds to 
simulation E with star formation recipe SF1 (constant timescale). 
The main difference is the slope of the  $\Sigma_{\rm SFR} - \Sigma_\H2$
relation.  For the constant time scale recipe SF1 the slope is shallower
and is close to unity. Note, however, that the slope of the 
$\Sigma_{\rm SFR} - \Sigma_{\rm gas}$ relation does not change as dramatically
and in fact is fairly consistent with the best fit relation of \citet{sfr:k98a}. 
This is because this slope is controlled both by the dependence of 
local star formation efficiency on density {\it and} on the dependence
of mass fraction of star forming regions on $\Sigma_{\rm gas}$ \citep{sims:k03}.

\section{Discussion and Conclusions}\label{disc}

Specifics of star formation modeling in cosmological simulations of galaxy 
formation have critical impact on the properties of resulting galaxies, 
such as their star formation histories (and hence luminosity and colors) and
morphology. Although significant successes have been achieved using such
high-resolution simulations \citep[c.f.,][for a comprehensive review]{sims:mgk08},
many challenges remain. In particular, the two likely related 
challenges are the low efficiency 
of star formation in low mass systems and the prevalence of thin
disks among galaxies. So far, the answer for these challenges was to 
devise efficient schemes of stellar energy feedback, which is capable of driving outflows \citep[e.g.,][]{springel_hernquist03,stinson_etal06,sims:ck07}.
However, at least part of the solution may be due to the inherently low efficiency
of gas conversion into stars \citep[e.g.,][]{sfr:km05}. Indeed, both nearby and high-redshift
galaxies provide a variety of clues suggesting that gas conversion into stars 
in low-mass, low-metallicity galaxies is very inefficient. To explore
the effects of such inefficiency on galaxy evolution, a more realistic way
of identifying and treating star forming regions is needed. In particular, the
global efficiency with which a galaxy converts gas into stars depends
on its ability to convert a significant fraction of its gas mass into
fully molecular form, within which conditions for star formation can
be realized. The latter depends on the small-scale gas properties and
local interstellar radiation field. These dependencies are not captured
in the standard recipes of star formation. 

In this paper we have presented a model for molecular hydrogen formation for
cosmological galaxy formation simulations designed to follow the
transition from the atomic to molecular phase on the scales of tens of
parsecs. The model is applicable in simulations, in which individual
star forming regions -- the 
giant molecular complexes -- can be identified and their mean internal density 
estimated reliably, even if internal structure is not resolved. We present
a number of tests of the model and illustrate its effects on the
global correlation 
between star formation rate and gas surface density. 

The model shows that the transition from atomic to fully molecular
phase depends primarily on the metallicity (which we assume is
directly related to the dust abundance) and clumpiness of the
interstellar medium.  The clumpiness simply boosts the formation rate
of molecular hydrogen, while dust both serves as a catalyst of $\H2$
formation and as additional shielding from dissociating UV
radiation. The upshot is that it is difficult to form fully-shielded
giant molecular clouds, while gas metallicity is low. However, once
the gas is enriched (say to $Z\sim 0.01-0.1Z_{\odot}$), the subsequent
star formation and enrichment can proceed at a much faster rate.  This
may keep star formation efficiency in the low-mass, low-metallicity
progenitors of galaxies very low for a certain period of time. One can
think of this as a ``feedback'' mechanism because it reduces
star formation in low metallicity galaxies, but also speeds up star
formation once conditions for sustained conversion of a large fraction
of gas into fully molecular form are realized.

Molecular fractions in the low-metallicity environments of the nearby
dwarf and low surface brightness galaxies are indeed very small 
\citep[][]{matthews_etal05,das_etal06}.
Recent observational evidence indicates that the fraction of star
forming molecular gas is also small in high-redshift galaxies
\citep[e.g.,][]{tumlinson_etal07,wolfe_chen06}. Dependence of the density
at which transition from atomic to fully molecular phase occurs on metallicity
is indeed observed both in direct measurements for nearby galaxies (see Fig.~\ref{nfH2_Z_col_C6})
and in the maximum HI column densities of DLAs of different metallicity
at high redshifts \citep{schaye01} and in the surface density of gas at which surface
density of HI saturates in nearby galaxies \citep{krumholz_etal08}. 

The UV radiation field is also important in maintaining the cold neutral 
medium in atomic form \citep{robertson_kravtsov08}. Our results show, however,
that beyond certain flux level the effect of UV radiation on the ability
of ISM to form molecular clouds saturates. 

We show that the global Kennicutt-Schmidt relation between gas and
star formation surface densities is also affected by the local
processes controlling conversion of atomic gas into molecular. Our
results are qualitatively consistent with those of
\citet{robertson_kravtsov08} in that star formation rate dependence on
the total gas density can vary depending on the local conditions in
the interstellar medium. These results are also consistent with
existing detailed studies of the Kennicutt-Schmidt relation in nearby
galaxies with lower metallicities and surface
densities \citep[e.g.,][Wyder et al. 2008, in
preparation]{boissier_etal03,heyer_etal04}.

The most recent comprehensive study of the KS relation in the THINGS
galaxy sample by \citet{bigiel_etal08} shows the KS relation
qualitatively similar with our results presented in
Figures~\ref{NSFLAW_l3_Z_C6} - \ref{NSFLAW_l3_sfr_C6}: $\Sigma_{\rm
SFR}$ dependence on $\Sigma_{\rm gas}$ is steep at low gas surface
densities, but becomes shallow at larger $\Sigma_{\rm gas}$
values. Interestingly, they find evidence that the slope $n$ of the
KS relation at $\Sigma_{\rm gas}\sim 10-100{\rm\, M_{\odot}}$ is
$n\approx 1\pm0.2$, shallower than the canonical value of $n\approx 1.4$.
The relation at higher surface densities in star burst galaxies
steepens \citep[see Fig.~15 of][]{bigiel_etal08}.  In the context of our calculations, these results can be
interpreted as the systematic change of the internal density of molecular
clouds. In our star formation recipes the slope of the large-scale relation
depends on the assumed dependence of star formation time scale, $\tau_{\rm sf}$, 
on the local internal density on the scale of star forming regions, as 
shown in Figure~\ref{NSFLAW_l3_sfr_C6}. The steepening of the observed
KS relation at high surface densities can thus be interpreted as due 
to systematic increase of the internal density of molecular clouds 
at high surface densities, as can be expected in the high-pressure environments
of starburst galaxies \citep[e.g.][]{downes_solomon98}. 

Another prediction of our calculations is that the transition at low gas surface
densities from a steep to a shallow relation should depend on metallicity. 
It would thus be interesting to also explore such metallicity dependence 
in observations. As a first step, one can consider comparison of
 the KS relation in lower-metallicity dwarf galaxies to the low-$\Sigma_{\rm gas}$
regions in the outskirts of higher-metallicity spirals by \citet[][see their Fig.~12]{bigiel_etal08}. 
Although the KS relation is steep in both of these regimes, the relation 
for the outskirts of spirals is shifted somewhat to lower $\Sigma_{\rm gas}$, 
as could be expected from the metallicity dependence. 

The above comparisons illustrate that a star formation model of the kind presented in
this paper \citep[see also][]{robertson_kravtsov08} can be very useful
in interpreting the detailed features of the observed KS relation.
Conversely, one can use the detailed observations of the KS relation in 
different regimes (different $\Sigma_{\rm gas}$, different metallicities, etc.)
to constrain and tune model parameters. This, in turn, may significantly improve 
fidelity of star formation modeling in simulations. We plan to present such comparisons
as well as applications of our model in fully self-consistent cosmological
simulations of galaxy formation in a forthcoming work.

\acknowledgements 

We thank Brant Robertson and Adam Leroy for comments on the manuscript. 
This work was supported in part by the DOE at Fermilab, by the HST Theory grant
HST-AR-10283.01, by the NSF grants AST-0239759,
AST-0507666, and AST-0708154 and by NASA grant NAG5-13274. The research was also 
partially supported by the Kavli Institute for Cosmological Physics at
the University of Chicago through grant NSF PHY-0551142 and an endowment
from the Kavli Foundation.  Part of this work was carried out at the 
Jet Propulsion Laboratory, California Institute of Technology, under a 
contract with the National Aeronautics and Space Administration.  
Supercomputer simulations were run on the
IBM P690 array at the National Center for Supercomputing Applications
(under grant AST-020018N) and on the
Joint Fermilab - KICP Supercomputing Cluster, supported by grants from
Fermilab, Kavli Institute for Cosmological Physics, and the
University of Chicago. This work made extensive use of the NASA
Astrophysics Data System and {\tt arXiv.org} preprint server. 

\appendix

\subsection{Numerical Considerations}

Ideally, in a simulation with infinite spatial resolution, our model
should work as designed. In practice, however, the spatial resolution
of a simulation is finite. In addition, numerical solutions of partial
differential equations always contain truncation errors. In our case,
these errors lead to a small, but non-negligible amount of numerical
heating and advection that biases the $\H2$ formation model.

\begin{figure}
\plotone{\figname{1zone.ps}{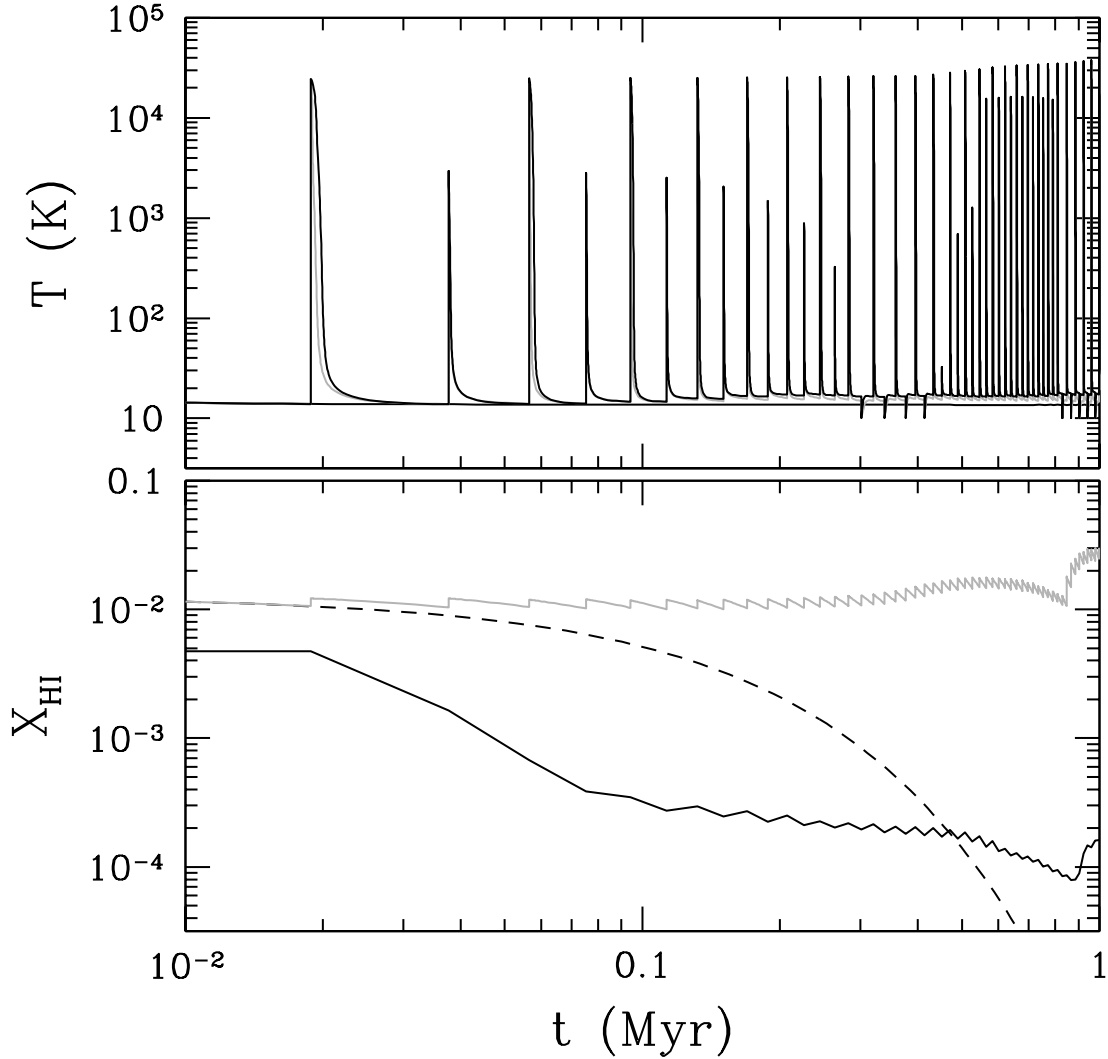}}
\caption{Evolution of the gas temperature (top) and $\HI$ fraction
  (bottom) for a single, fully molecular cell in our fiducial run. The
  black dashed line shows a test calculation when the cell is
  evolved in isolation at constant gas density and constant radiation
  field (without any hydrodynamic effects). The light gray solid
  line is the evolution of the same cell in a full cosmological
  simulation - notice the numerical heating and advection, appearing
  as jumps in temperature and neutral hydrogen fraction at regular
  intervals corresponding to individual hydrodynamic time steps. The
  solid black line is for the same test with the numerical correction
  (\ref{eq:numcor}).} 
\label{fig:1zone}
\end{figure}
An example of such diffusion is shown in Figure \ref{fig:1zone},
where we compare an ideal, single zone (i.e.\ without any hydrodynamic
effects) evolution of a typical cell inside molecular clouds (i.e.\
with molecular fraction above 99\% - shown as black dashed lines) with
the evolution of this cell in a real cosmological simulation. Because
our resolution is finite, gas motions across individual cells are not
fully resolved; in a Riemann code the truncation errors from such
incomplete resolution lead to cells being heated at each time
step. This numerical heating is not too serious, since the gas cooling
time is very short; however, a more serious artifact is the numerical
advection of atomic gas inside molecular clouds. The gray line on the
bottom panel shows the evolution of the atomic hydrogen fraction in
that cell - at each hydrodynamic time step there is a small (about
10\%) increase in the atomic hydrogen fraction. This amount is not
large, but it happens at every time step, and so, effectively, there
is a numerical flux of atomic hydrogen into molecular clouds, with the
corresponding opposite flux of molecular hydrogen out of molecular
gas.

We emphasize that this is a numerical artifact: the transition between
molecular and atomic phases is so sharp that few numerical schemes
are able to treat it perfectly. As a numerical artifact, this effect
is actually rather small - 10\% jump in $f_\HI\sim0.01$ corresponds to
the relative numerical diffusion of only about $10^{-3}$ per time
step. This effect is only important because the atomic hydrogen
fraction inside molecular clouds is of the same order, about
$10^{-3}$, or even less.

In order to mitigate this numerical advection, we adopt the following,
entirely empirical approach. The numerical advection can be thought of
as an additive increase in the atomic hydrogen fraction per time step,
$$
  X_{\HI}^{\rm Num} = X_{\HI}^{\rm True} + \Delta X_{\HI}.
$$
If we multiply the ``numerical'' value $X_{\HI}^{\rm Num}$ at the end
of each hydrodynamic time step by the correction factor
$$
  f_{\rm Num} = \left(1 + \frac{\Delta X_{\HI}}{X_{\HI}^{\rm
  True}}\right)^{-1},
$$
then we would remove the numerical advection
completely. Unfortunately, we do not have any way to measure both
$X_{\HI}^{\rm True}$ and $\Delta X_{\HI}$ in a self-consistent way
(for example, keeping track of $X_{\HI}$ without hydrodynamic
advection would violate Galilean invariance). Because the numerical
advection per time step is not large, we can use $X_{\HI}^{\rm Num}$
(the value we get in the simulation) instead of $X_{\HI}^{\rm True}$
in the expression for $f_{\rm Num}$, but the factor $\Delta X_{\HI}$
needs to be deduced heuristically. 

We choose to define $\Delta X_{\HI}$ as
\begin{equation}
  \Delta X_{\HI} \equiv \alpha \sigma_{\rm cell} \frac{\Delta
  t}{\Delta x} X_{\rm H},
  \label{eq:deltaxfac}
\end{equation}
where $\alpha$ is a numerical coefficient of the order of unity,
$\sigma_{\rm cell}$ is the gas velocity dispersion at the cell scale,
computed in each cell $(i,j,k)$ from its six neighbors,
\begin{equation}
  \sigma_{\rm cell}^2(i,j,k) \equiv \frac{1}{6} \left[
  \left(\vec{v}_{i+1,j,k}-\vec{v}_{i,j,k}\right)^2 + 
  \left(\vec{v}_{i-1,j,k}-\vec{v}_{i,j,k}\right)^2 + 
  \left(\vec{v}_{i,j+1,k}-\vec{v}_{i,j,k}\right)^2 + ... \right],
\end{equation}
and $\Delta t$ and $\Delta x$ are the numerical time step and the cell
size (which is, in fact, different for different cells on an
adaptively refined mesh in our simulations). The factor $X_{\rm H}$
encapsulates our assumption that the gas is mostly atomic and neutral
just outside molecular clouds.

Equation (\ref{eq:deltaxfac}) has the correct properties for an
expression describing numerical advection. First, it vanishes in the
limit of $\Delta t \rightarrow 0$. We have actually verified, by
running test simulations with reduced time steps, that the numerical
advection effect shown in Figure\ \ref{fig:1zone} scales linearly with
the time step $\Delta t$. Second, equation \ref{eq:deltaxfac} is
Galilean invariant and depends
on the local variation in the gas velocity, i.e.\ on the quantity that
determines hydrodynamic advection (beyond trivial uniform
translation). Third, $\Delta X_{\HI}$ does not actually diverge in the
limit $\Delta x \rightarrow 0$, because for a fully resolved flow
$$
  \sigma_{\rm cell} = \frac{1}{3}\left(\sum_{i,j}
  \frac{\partial v^i}{\partial x^j}
  \frac{\partial v^i}{\partial x^j}
  \right)^{1/2} \Delta x.
$$
Thus, our heuristic numerical advection correction factor becomes
\begin{equation}
  f_{\rm Num} \approx \left(1 + \alpha \sigma_{\rm cell} 
  \frac{\Delta t}{\Delta x}
  \frac{X_{\rm H}}{X_{\HI}}\right)^{-1}.
  \label{eq:numcor}
\end{equation}
This form for the numerical correction is also rather insensitive to
the specific choice of the coefficient $\alpha$: varying $\alpha$ from
0.5 to 2 changes atomic hydrogen fractions inside molecular clouds and
star formation rates by less than their natural scatter.

\begin{figure*}[t]
\plotone{\figname{fnum.ps}{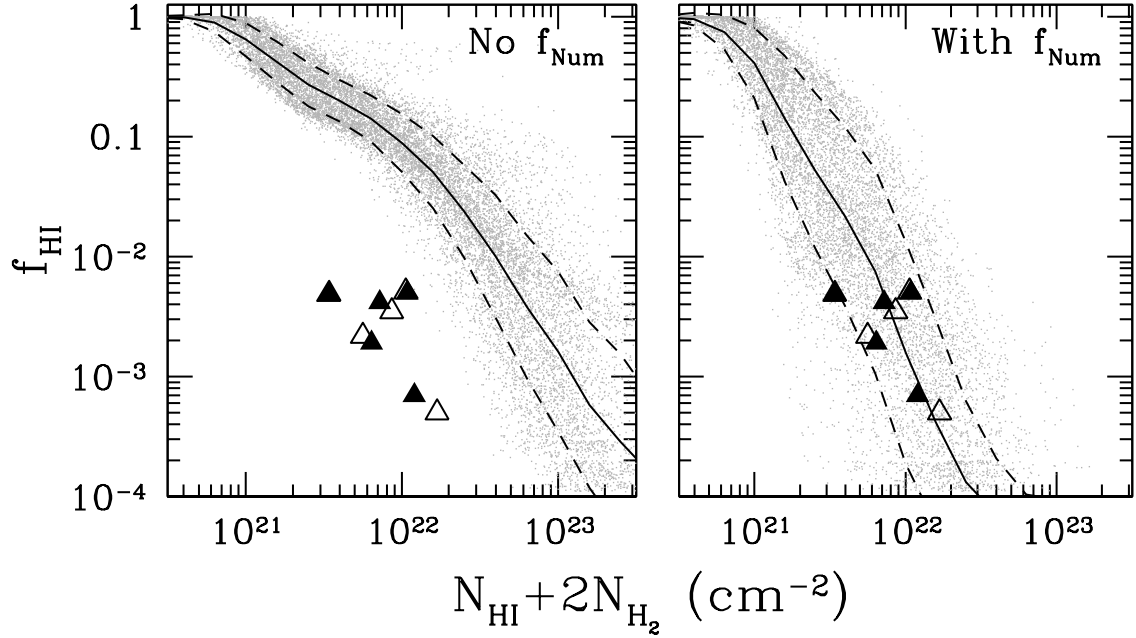}}
\caption{Atomic hydrogen fraction as a function of the total neutral
  hydrogen column density for two test runs: a run without numerical
  correction (\ref{eq:numcor}; left) and a run with numerical
  correction included (right). Gray points show individual 
  simulation cells, while the black solid and dashed line show the
  mean and $\pm1\sigma$. Black and open triangles are observational
  measurements from \citet{GL05}.}
\label{fig:fnum}
\end{figure*}

The correction factor (\ref{eq:numcor}), being heuristic, does not
completely correct for numerical advection on a cell-by-cell basis; it
is valid only in a statistical sense. As Figure\ \ref{fig:1zone} shows,
evolution of the abundance in a test cell with the correction factor
included does {\it not\/} match a single zone calculation (although, a
single zone calculation excludes not only unphysical numerical
advection, but physical real advection as well, and should not be
taken as a correct solution). The only justification for the factor
(\ref{eq:numcor}) is Figure\ \ref{fig:fnum}, which shows a comparison of
the atomic hydrogen fractions that we find in our simulations with the
observational points from \citet{GL05}. Without the correction
(\ref{eq:numcor}) the observational points cannot be reproduced, while
with the correction we get a good match to the observational constraints.

\bibliographystyle{apj}
\bibliography{ms,sims,sfr,h2}

\end{document}